# Mechanical and suture-holding properties of a UV-cured atelocollagen membrane with varied crosslinked architecture

Ruya Zhang,[1] Charles Brooker,[1,2] Laura L.E. Whitehouse[1], Neil H. Thomson,[1,3,4] David Wood,[1,4] Giuseppe Tronci[1,2,*]

[1]School of Dentistry, St. James's University Hospital, University of Leeds, LS9 7TF, UK

[2]Clothworkers' Centre for Textile Materials Innovation for Healthcare (CCTMIH), School of Design, University of Leeds, LS2 9JT, UK

[3]School of Physics and Astronomy, University of Leeds, Leeds, LS2 9JT, UK

[4]Bragg Centre for Materials Research, University of Leeds, LS2 9JT, UK

**Abstract**

The mechanical competence and suturing ability of collagen-based membranes are paramount in Guided Bone Regeneration (GBR) therapy, to ensure damage-free implantation, fixation and space maintenance *in vivo*. However, contact with the biological medium can induce swelling of collagen molecules, yielding risks of membrane sinking into the bone defect, early loss of barrier function, and irreversibly compromised clinical outcomes. To address these challenges, this study investigates the effect of the crosslinked network architecture on both mechanical and suture-holding properties of a new atelocollagen (AC) membrane. UV-cured networks were obtained via either single functionalisation of AC with 4-vinylbenzyl chloride (4VBC) or sequential functionalisation of AC with both 4VBC and methacrylic anhydride (MA). The wet-state compression modulus ($E_c$) and swelling ratio ($SR$) were significantly affected by the UV-cured network architecture, leading up to a three-fold reduction in

[*] Corresponding author: g.tronci@leeds.ac.uk

*SR* and about two-fold increase in $E_c$ in the sequentially functionalised, compared to the single-functionalised, samples. Electron microscopy, dimensional analysis and compression testing revealed the direct impact of the ethanol series dehydration process on membrane microstructure, yielding densification of the freshly synthesised porous samples and a pore-free microstructure with increased $E_c$. Nanoindentation tests via spherical bead-probe Atomic Force Microscopy (AFM) confirmed an approximately two-fold increase in median (interquartile range) elastic modulus in the sequentially functionalised ($E_{AFM}$= 40 (13) kPa), with respect to single-functionalised ($E_{AFM}$= 15 (9) kPa), variants. Noteworthy, the single-functionalised, but not the sequentially functionalised, samples displayed higher suture retention strength (*SRS*= 28±2—35±10 N·mm$^{-1}$) in both the dry state and following 1 hour in Phosphate Buffered Saline (PBS), compared to Bio-Gide® (*SRS*: 6±1–14±2 N·mm$^{-1}$), while a significant decrease was measured after 24 hours in PBS (*SRS*= 1±1 N·mm$^{-1}$). These structure-property relationships confirm the key role played by the molecular architecture of covalently crosslinked collagen, aimed towards long-lasting resorbable membranes for predictable GBR therapy.

**Introduction**

GBR therapy is a surgical dental procedure aiming to support the regeneration of new bone in critical-sized bone defects (CSBDs), which may be formed consequent to periodontal disease, genetic conditions, trauma, or tumour resection. In GBR therapy, a barrier membrane is applied on the bone defect to provide an osteoconductive environment and prevent soft tissue growth [1].

The original non-resorbable barrier membrane made of expansion polytetrafluoroethylene (e-PTFE) and high-density polytetrafluoroethylene (d-PTFE) was first applied to heal maxillofacial defects in 1988 [2]. However, additional surgery is required to remove the non-resorbable membrane, exposing patients to risks of pain and other adverse events, including infection and wound complications. Consequently, resorbable membranes have gradually become the preferred choice for GBR therapy [3,4,5], stimulating growing research aiming to identify suitable degradable biomaterials.

Aliphatic polyesters have been initially investigated, due to their mechanical competence in a hydrated environment, hydrolytic degradability, and FDA approval for human use [6]. Poly($\varepsilon$-caprolactone) (PCL) has been integrated with other materials, e.g. hydroxyapatite and natural polymers, in various configurations to improve its slow degradation rate and accomplish osteogenic and antibacterial characteristics, as well as barrier function [7,8,9]. Polylactide has been employed in electrospinning to prepare composite fibrous membranes with integrated bioactivity, dimensional stability, and antibacterial properties [10,11]. Other research prototypes have been made of poly(lactic-co-glycolic acid) (PLGA), exploiting its adjustable physicochemical properties by variation in the block ratio of lactide to glycolic acid chain segments [12]. Multilayer PLGA-based membrane prototypes have also been developed with varying ratios of HA and β-tricalcium phosphate (TCP) to support osteoblast attachment, proliferation, and migration, while preventing fibroblast infiltration [13].

Other than aliphatic polyesters, type I collagen has been employed as a building block of commercial resorbable membranes, e.g. Bio-Gide® (Geistlich), BioMend Extend® (Zimmer), and Cytoplast® RTM (Osteogenics), due to its affinity for biomolecules as well as for its organisational and macromolecular features similar to the natural extracellular matrix (ECM) [14]. Most importantly, the enzymatic degradability and amino acidic composition of type I collagen prevent the generation of acidic, potentially toxic degradation products, which are formed during polyester hydrolysis and which are known to reduce bone regeneration, often leading to additional surgical interventions and longer treatment times [15]. Type I collagen is typically extracted from human, bovine, or porcine tissues, and can present minimal antigenicity following the removal of telopeptides [16,17,18], yielding atelocollagen (AC). However, collagen-based resorbable membranes used in GBR therapy still face performance limitations, particularly concerning their mechanical properties, space maintenance, and rapid degradation, especially in non-crosslinked variants [19,20]. Due to its highly hydrophilic nature and cleavage of natural crosslinks, type I collagen *ex vivo* often undergoes uncontrollable swelling in an aqueous environment [21,22,23], generating risks of dimensional instability and membrane collapse in the bone defect. Furthermore, the brittleness of dry collagen can limit clinical handling, defect-free implantation, and fixation to the surrounding tissues.

To address the aforementioned challenges, blends of type I collagen and natural polymers, i.e. chitosan [24,25] and cellulose [26], have been pursued in single or bilayer configurations, together with physical and chemical crosslinking methods, including UV irradiation, dehydrothermal treatment and carbodiimide chemistry. Promising results were observed *in vitro*, while long-term studies *in vivo* are required to assess GBR functionality during bone regeneration. Therefore, the design of long-lasting resorbable membranes enabling enhanced mechanical properties and space maintenance is still critical for successful GBR therapy.

This work investigates the bespoke covalent functionalisation of type I AC as a potential strategy to develop UV-cured resorbable membranes with enhanced mechanical and suture-holding properties for predictable GBR therapy. Our previous investigations on the single functionalisation of AC with 4VBC yielded a UV-cured network with wound healing capability, rapid gelation kinetics and remarkably high swelling *and* compression properties [27, 28, 29]. More recently, the sequential functionalisation of AC with 4VBC and MA generated a UV-cured system with competitive fibroblast barrier function, enhanced proteolytic stability *in vitro* and space maintenance capability following one month *in vivo* [30]. Leveraging the appealing and complementary features of the above UV-cured systems, we studied the effect of the crosslinked network architecture deriving from either single or sequential AC functionalisation on microstructure, mechanical competence, swellability and suturing ability, given the major impact of these characteristics on GBR functionality. The effect of post-synthesis sample drying was also investigated as an additional experimental dimension to adjust micro- and macroscale properties for GBR therapy. Comprehensive physicochemical characterisation was carried out by electron microscopy, compression and suture pull-out tests, AFM nanoindentation, and gravimetric analysis, to quantify material characteristics in the dry and near-physiological conditions, laying the foundations for new structure-property relationships and the design of future resorbable membranes.

## 2. Materials and methods

### 2.1 Materials

A solution of type I AC (6 mg·ml$^{-1}$) in 10 mM hydrochloric acid (HCl) was purchased from Collagen Solutions PLC (Glasgow, UK). Agarose (A2790-100G, Sigma-Aldrich), 4-vinylbenzyl chloride (4VBC), methacrylic anhydride (MA) and triethylamine (TEA) were purchased from Sigma-Aldrich (UK). 2-Hydroxy-4′-(2-hydroxyethoxy)-2-methylpropiophenone (I2959), the violet braided absorbable suture and Bio-guide® (25×25 mm) were purchased from Fluorochem Limited (Glossop, UK), ETHICON® (Coated VICRYL™, REF: W9105) and Geistlich (UK), respectively. Absolute ethanol was purchased from VWR Internationals. All other chemicals were supplied by Sigma-Aldrich.

**2.2 Single and sequential functionalisation of collagen**

The synthesis of 4VBC-functionalised AC was carried out as previously reported [31]. A solution of medical grade type I bovine AC (6 mg·ml$^{-1}$) was diluted to a concentration of 3 mg·ml$^{-1}$ in 10 mM hydrochloric acid (HCl) and neutralised to pH 7.5 by addition of 0.1 M NaOH. Polysorbate 20 (PS-20) was added at a concentration of 1 wt.% (with respect to the weight of the solution prior to neutralisation), followed by the introduction of 4VBC and TEA with a 25 molar ratio with respect to the molar content of AC lysines ([4VBC]·[Lys]$^{-1}$ =25; [4VBC] = [TEA]). Following a 24-hour reaction, the mixture was precipitated in a 10-volume excess of absolute ethanol and incubated for 8 hours. The solution was subsequently centrifuged, and the pellet recovered and air-dried. The resulting product was coded as *4VBC*.

For the synthesis of the sequentially functionalised AC [30], the 4VBC-functionalised product was then dissolved in 10 mM HCl (3 mg·ml$^{-1}$) under magnetic stirring at room temperature. Following solution neutralisation (pH 7.5), MA and TEA were added ([MA]·[Lys]$^{-1}$ =25; [MA]= [TEA]) to the above solution and the reaction was carried out for 24 hours. The product was recovered and air-dried as previously discussed, and coded *4VBC-MA*.

**2.3 Preparation of dental membranes**

The photoinitiator 2-Hydroxy-4'-(2-hydroxyethoxy)-2-methylpropiophenone (I2959) was dissolved (1 wt.%) in a 10 mM HCl solution at 45 °C in the dark for one hour. The respective solution was then cooled down to room temperature. Samples of either single- (4VBC) or sequentially functionalised (4VBC-MA) AC were dissolved (1.2 wt.%) in the I2959-supplemented solution via magnetic stirring at room temperature for at least 2 days. Solutions of functionalised AC were centrifuged at 3000 rpm for 5 min to remove air bubbles, cast onto a 24-well plate (0.6-0.8 g in each well) and exposed to UV (Chromato-Vue C-71, Analytik Jena, Upland, CA, USA) for 15 min on both top and bottom sides. The freshly synthesised UV-cured hydrogels were coded as either *F-4VBC\** or *F-4VBC-MA\**, where *F* and * indicate the freshly synthesised and crosslinked states, respectively. The UV-cured samples were rinsed in distilled water and dehydrated in a series of distilled water-ethanol mixtures (0, 20, 40, 60, 80, 100 wt.% EtOH) prior to air-drying. The UV-cured networks that were dried according to the aforementioned ethanol series dehydration process were coded as either *4VBC\** or *4VBC-MA\**, where * has the same meaning as discussed above. Alternatively, freshly synthesised samples were frozen at -20 °C (12 hours) and freeze-dried at -54°C for 48 hours. Agarose controls were obtained by dissolving agarose (3-12 wt.%) in distilled water at 70 °C and cast into either 96-well plates or Petri dishes.

## 2.4 Scanning electron microscopy (SEM)

Freshly synthesised, ethanol dehydrated and freeze-dried samples of UV-cured AC and cast agarose were observed via variable pressure SEM (Hitachi S-3400N VP, 60-70Pa) and ESED, BSECOMP and BSE3D detectors with cool stage control (Model: LT3299; Temperature: 25°C). All agarose controls were imaged following either solution casting and equilibration to room temperature or the ethanol series dehydration process.

## 2.5 Swelling tests

The dehydrated 4VBC* and 4VBC-MA* samples (n=3) were individually weighed ($m_d$) prior to room temperature incubation in either distilled water, PBS (10 mM, pH 7.4), or saline solution (0.9 w/v% NaCl). At specific incubation time points, the samples were collected and paper blotted prior to weighing ($m_s$). The swelling ratio (SR) was calculated as weight percentage (wt.%) according to Equation 1:

$$SR = \frac{m_s - m_d}{m_d} \times 100 \qquad \text{(Eq. 1)}$$

The samples equilibrated in PBS were coded as *R-4VBC\** and *R-4VBC-MA\**, where *R* indicates the rehydrated state of the samples and * has the same meaning as previously discussed. The samples equilibrated in distilled water were coded as *H-4VBC\** and *H-4VBC-MA\**, where *H* specifies that the rehydration solvent is distilled water.

**2.6 Compression tests**

Ethanol-dehydrated samples of 4VBC* and 4VBC-MA* were equilibrated in PBS aiming to simulate near-physiological hydrated conditions. Compression testing (n=3, Ø: 6 mm) was performed at room temperature up to sample failure, using a 250 N load cell and a compression rate of 3 mm·min$^{-1}$ (Bose EnduraTEC ELF3200). The compression modulus was quantified via linear fitting in the strain range associated with 10–15% of maximal stress [31].

**2.7 Atomic force microscopy**

Ethanol-dehydrated UV-cured networks were rehydrated in PBS for 24 hours, whereby PBS was selected aiming to simulate near-physiological hydrated conditions. The PBS-equilibrated samples were cut to a height of 2 mm using a razor blade, prior to gluing to a circular glass coverslip with a tissue adhesive (3M Vetbond™, No.1469SB). Agarose controls (3-12 wt.% agarose) were prepared by solution casting, followed by equilibration to room temperature and 24-hour incubation in PBS. The construct of either AC or agarose samples was mounted on a liquid stage kit (BioHeater™, Asylum

Research) and placed on the sample stage of an MFP-3D AFM (Asylum Research, Oxford Instruments, Santa Barbara, CA, USA). Approximately 100 µl of PBS solution was placed on top of the sample surface to keep it fully hydrated. AFM nanoindentation measurements were carried out using a high-density carbon, pre-calibrated spherical bead tip with a controlled radius of 1500±10 nm, integrated onto a silicon cantilever (*l*: 450 µm, *w*: 50 µm, *h*: 2 µm) with a nominal spring constant of 0.2 N·m$^{-1}$ and a resonant frequency of 13 kHz (biosphere B1500-CONT; nanotools GmbH, supplied by Apex Probes Ltd, Bracknell, UK). Contact mode was selected, and the detector sensitivity was calibrated by taking a force curve on the hard glass slide surface. The cantilever spring constant was then determined using the thermal tune method with the tip off the surface [32]. This procedure was first carried out in air and then in the liquid environment to ensure validation of spring constant determination, before sample measurement. Force volume (FV) measurements were performed at relatively low spatial resolution in five different areas in the central region of each sample. Each FV map (FVM) was a 10×10 µm array, at 1 µm spatial XY separation corresponding to 100 force curves (FC) per map. The ramp size was set between 1 to 3 µm at a frequency of 1 to 3 Hz, with the maximum loading force in the range of 10 to 50 nN (dependent on the sample modulus) adjusted to give a typical mean indentation depth of ~500 nm. The tip-sample adhesion force was relatively low (< 20%) compared to the typical maximum indentation loading forces (up to 50 nN), so the elastic modulus ($E_{AFM}$) was determined by fitting the indentation curve using the Hertz contact mechanics model for a spherical tip:

$$F = \frac{4}{3}\left(\frac{E_{AFM}}{1-\nu^2}\right)\sqrt{R} * \delta^{\frac{3}{2}} \quad \text{(Eq. 2)}$$

where *F* is the loading force, *v* is the Poisson's ratio, which was set at 0.5 as a typical value for biopolymer-based samples [28]; *R* is the radius of the tip and $\delta$ is the indentation.

Parameter maps were extracted from the FV arrays for sample height and $E_{AFM}$ at the maximum indentation, as well as the adhesion force ($F_A$), i.e. the maximum negative force on bead retraction from the sample. To check that the surface topography of the sample was not influencing either the $E_{AFM}$ or $F_A$, image cross-correlation was performed between the height maps and either $E_{AFM}$ or $F_A$ using the Image_Correlator.class plugin in ImageJ (v. 1.54). This returns a cross-correlation coefficient ($r_{xy}$) value between +/-1, so that the closer the value of $r_{xy}$ is to +1, the more closely the two images are correlated.

**2.8 Suture pull-out measurements**

Suture pull-out measurements were carried out with freeze-dried UV-cured AC samples as well as with AC samples incubated in PBS for either 1 hour or 24 hours, alongside Bio-Guide® as the clinical gold standard. The resulting samples of 4VBC* and 4VBC-MA* were cut into square shapes and sutured 2 mm below the top edge using 5-0 absorbable sutures and round needles. The bottom edge of the sutured sample was firmly adhered to a thick piece of paper by the application of a commercial super glue (Super Glue®, No.15187), and then placed on a tensile testing machine (INSTRON 3365), prior to fixation of the top suture [33]. Tensile tests were performed with a strain rate of 0.02 mm·s$^{-1}$ until the membrane was fully broken. Three replicates were used for each sample group and the resulting force was normalised by the sample height. The *SRS* was quantified as the height-normalised maximal force following catastrophic failure and suture pull-out, while the break starting strength (*BSS*) was recorded as the height-normalised force at the earliest failure of the specimen [34].

**2.9 Statistical analysis**

Statistical analysis was carried out using Origin (version 2023b Academic). Data normality was confirmed prior to the determination of statistical significance with either t-test or one-way ANOVA corrected with either the Bonferroni or Tukey HSD test. Normally distributed data were presented as mean ± standard deviation. When data normality was rejected, data were presented as median (interquartile range, IQR) and statistical significance determined via Kruskal-Wallis test.

**Results and discussion**

Collagen membranes with controlled structure-property relationships are key for predictable GBR therapy, aiming to accomplish bone regeneration, prevent soft tissue invasion, and avoid the need for a second surgery, which is otherwise required with non-resorbable membranes. The uncontrollable swelling in the biological environment and the water-induced plasticisation of collagen molecules contribute to the loss of mechanical competence in, and fixation *in situ* by, current collagen-based resorbable membranes. Two UV-cured prototypes made of either single- or sequentially functionalised AC were comprehensively investigated, aiming to establish the effect of molecular architecture and post-synthesis drying on microstructure, mechanical properties and suturing ability (Figure 1).

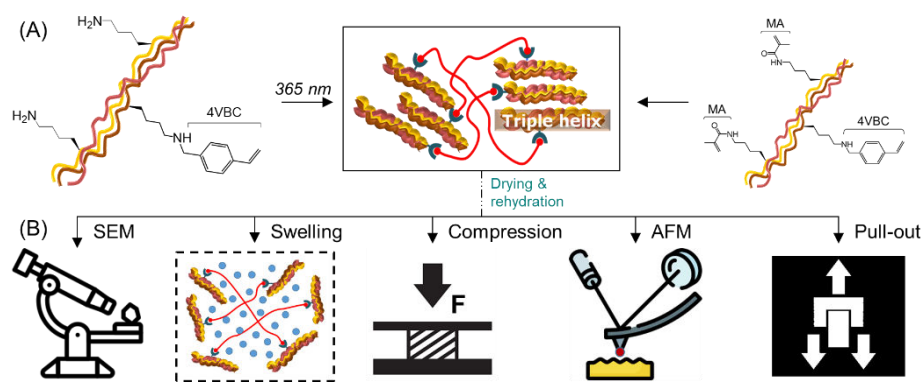

**Figure 1.** Research strategies adopted to assess GBR functionality. (A): Two UV-cured membranes were prepared from aqueous solutions of either single-

functionalised (4VBC) or sequentially functionalised (4VBC-MA) AC. (B): Following sample drying and rehydration in PBS, comprehensive physical characterisation was carried out via SEM, swelling and compression tests, AFM, and suture pull-out measurements.

**3.1 Effect of drying on membrane microstructure**

Membrane prototypes were obtained via UV-curing of photoactive aqueous solutions containing either single- or sequentially functionalised AC. In addition to the different type of photoactive residues grafted to AC, respective UV-cured networks were expected to display significantly different crosslink densities at the molecular scale, given that a degree of functionalisation of 21±3 mol.% or 97±2 mol.% was measured in samples of 4VBC and 4VBC-MA [30,31], respectively. Consequent to the UV-induced gelation of each functionalised AC solution, resulting hydrogel networks were dried according to two distinct drying protocols, i.e. ethanol series dehydration or freeze-drying, aiming to accomplish purification and removal of unreactive chemicals, on the one hand, and minimal alteration of the UV-cured microstructure, on the other hand.

Figure 2 depicts the typical SEM images of the UV-cured samples following synthesis and the ethanol series dehydration process, indicating a significant impact of the sample history on the sample microstructure. Freshly synthesised samples F-4VBC* and F-4VBC-MA* revealed the presence of pores (Figure 2A-D and Figure S1, Supp. Inf.), irrespective of whether single- or sequentially functionalised AC was solubilised prior to UV-curing, in line with previous reports [28,30]. Given that the same concentration of AC precursor was employed in both hydrogel-forming solutions, the detection of micropores in the resulting networks is attributed to the free volume in the covalent network [35], the afibrillar organisation of collagen molecules [36] during UV-curing and the presence of air bubbles in the hydrogel-forming solution [37]. This explanation was corroborated by the SEM images of agarose hydrogel controls prepared according to the same casting process (Figure S2, Supp. Inf.), whereby porous-like microstructures were evident at relatively low concentrations (1.5-3 wt.%), in line with previous reports [38,39] and comparably to the one employed for the UV-curing of the AC precursors (1.2 wt.%). Of note, this effect was less evident at higher concentrations of biopolymer, whereby the concentration-induced increase in solution viscosity makes it increasingly difficult to achieve a uniform dispersion, resulting in microstructural heterogeneities in the corresponding film [40].

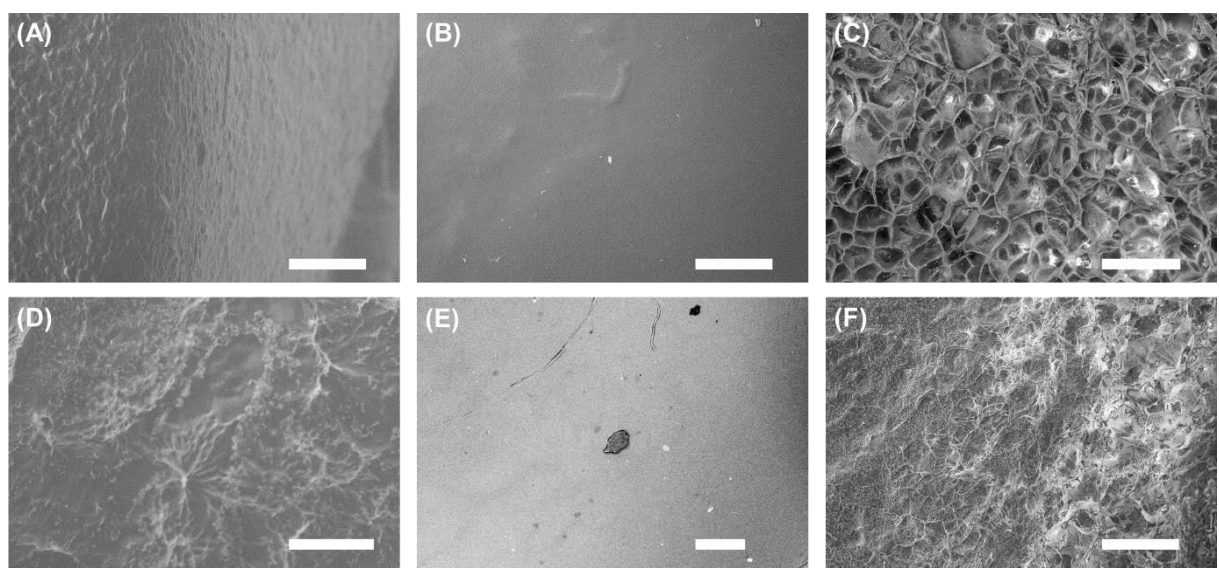

**Figure 2.** Cool-stage SEM images of UV-cured collagen samples 4VBC* (A-C) and 4VBC-MA* (D-F). Images were captured directly after UV-curing (A & D), following ethanol-induced sample dehydration and rehydration in PBS (10 mM, pH 7.4; B & E), or following freeze-drying of UV-cured samples (C & F). Scale bar ≈ 250 μm.

When the freshly prepared samples of F-4VBC* and F-4VBC-MA* were dehydrated according to the ethanol series dehydration process, a pore-free microstructure was observed, whereby previous micropores were lost (Figure 2B & E). The lack of detectable internal micropores in ethanol-dehydrated samples was associated with significant macroscale shrinking and volume reduction ($R_V$ ~ 95 %) with respect to the freshly prepared samples (Table 1). The sample height decreased from 4.13±0.76 mm and 6.19±1.35 mm in F-4VBC* and F-4VBC-MA* (i.e. freshly synthesised samples) to 0.20±0.02 mm and 0.34±0.04 mm in 4VBC* and 4VBC-MA* (i.e. the corresponding ethanol-dehydrated samples), respectively. Following incubation in water, samples H-4VBC-MA* ($h$= 3.91±0.75 mm; $V$= 111.54±21.17 mm$^3$) and H-4VBC-MA* ($h$= 5.81±0.98 mm; $V$= 156.47±25.11 mm$^3$) recovered their freshly synthesised shape, though with a loss of roughly 6% of their original volume (Table 1). This observation agrees with the aforementioned suppression of internal micropores detected in the ethanol rehydrated samples, in contrast to the case of the freshly synthesised variants.

**Table 1.** Dimensions of freshly synthesised (F-4VBC* and F-4VBC-MA*), ethanol-dehydrated (4VBC* and 4VBC-MA*) and water-rehydrated (H-4VBC* and H-4VBC-MA*) collagen membranes (n=3). $h$: Height; Ø: diameter; $V$: volume: $R_V$: reduction in volume measured in the water-rehydrated samples compared to the corresponding freshly synthesised samples. N/A: not applicable.

| Sample ID | $h$ /mm | Ø /mm | $V$ /mm$^3$ | $R_V$ /% |
|---|---|---|---|---|
| F-4VBC* | 4.13±0.76 | 6.06±0.05 | 118.89±21.80 | N/A |
| F-4VBC-MA* | 6.19±1.35 | 5.89±0.04 | 168.08±34.27 | N/A |
| 4VBC* | 0.20±0.02 | 6.01±0.05 | 5.68±0.56 | 95.2±0.4 |
| 4VBC-MA* | 0.34±0.04 | 5.81±0.09 | 9.00±1.03 | 94.6±0.5 |
| H-4VBC* | 3.91±0.75 | 6.03±0.04 | 111.54±21.17 | 6.2±0.6 |
| H-4VBC-MA* | 5.81±0.98 | 5.87±0.03 | 156.47±25.11 | 6.4±3.8 |

Given that comparable results were obtained when the same process was applied to the agarose controls (Figure S3, Supp. Inf.), the lack of internal micropores observed in samples of 4VBC* and 4VBC-MA* can be explained by the fact that the dehydration of the freshly synthesised samples induces densification of the membrane microstructure, consequent to solvent exchange, water removal and the development of secondary interactions between biopolymer molecules [41]. On the other hand, a preserved porous microstructure was still detectable when the freshly synthesised AC networks were freeze-dried (Figure 2C & F), in line with the removal of water molecules through the sublimation of ice formed in the frozen samples.

Overall, the above results indicate that it is possible to adjust the microstructure of the UV-collagen networks through the specific drying procedure adopted post-synthesis, so that either bulky or porous materials can be obtained. Beneficially, the ethanol-induced dehydration route promotes densification and suppression of the native hydrogel architecture and the creation of a GBR prototype with inherent barrier functionality. This is particularly appealing for preventing soft tissue infiltration into the bone defect following implantation of the GBR membrane *in vivo*. For these reasons, this sequential drying procedure was selected for the subsequent characterisation of the UV-cured prototypes. The aforementioned dehydration process through sequential ethanol concentration series can also enable the extraction of unreacted chemicals, e.g. photoinitiator or acidic residues, through diffusion of these into the ethanol-supplemented mixtures, which is key to ensuring prototype purification and minimising risks of cytotoxicity [30]. On the other hand, freeze-drying of the AC hydrogel networks enables solvent removal through sublimation, so that preservation of the original macroscopic shape and inner micropores generates structures suitable for tissue engineering applications.

## 3.2 Swelling properties in near-physiological environments

The water-induced swelling of collagen-based GBR membranes plays a major role in their clinical performance. A decreased swelling ratio (*SR*) is advantageous for wound closure and barrier function [42,43], while an uncontrolled water uptake can lead to membrane displacement, accelerated hydrolysis and poor bone quality formation. To address this point, the ethanol-dehydrated samples of both 4VBC* and 4VBC-MA*, alongside Bio-Gide®, were incubated in either distilled water, PBS or saline solution, as near-physiological environments *in vitro*, and respective swelling properties quantified gravimetrically.

Statistically different swelling profiles were observed between the two UV-cured AC membranes in all swelling media investigated (Figure 3A-B & D). Samples made of the single-functionalised AC precursor indicated the highest values of *SR* following 24-hour incubation in each aqueous medium (Figure 3D), corresponding to an increase of *SR* of up to 68 wt.% with respect to the UV-cured variant made of the sequentially functionalised AC. When comparing the *SR* across the three aqueous environments, the highest values were recorded in distilled water, followed by PBS, and saline solution, such that a statistical significance in *SR* was recorded between the former and latter medium (Figure 3D).

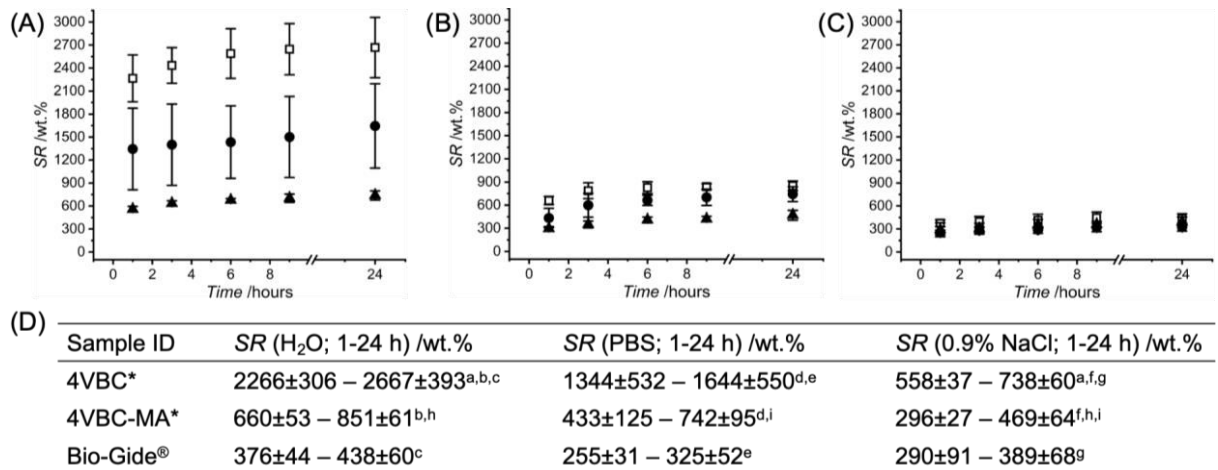

**Figure 3.** (A-C): Swelling ratio (*SR*; n=3) revealed by samples of 4VBC* (A), 4VBC-MA* (B) and Bio-Gide® (C) during a 24-hour incubation in distilled water (□), PBS (●), and saline solution (▲). (D): Extracted values of 1-hour *SR* and 24-hour *SR* in the aforementioned media. Data are presented as mean ± standard deviation. Same superscripts indicate statistically significant means of the 24-hour values of *SR* ($p < 0.05$, Bonferroni test).

The increased *SR* measured in distilled water compared to the one measured in either PBS or saline solution is attributed to the presence of salts in the latter media, which have been reported to generate reduced swelling capability in crosslinked collagen [44]. In particular, the presence of chloride ions is expected to inversely affect the water uptake in the network due to the electrostatic shielding effect of these ions and consequently decreased osmotic pressure between the covalent network and the aqueous solution [45,46].

When the swelling profile of Bio-Gide® was considered, insignificant differences were recorded with respect to samples of 4VBC-MA* following 24-hour incubation in either distilled water, PBS, or saline solution (Figure 3C-D). On the other hand, a significant decrease in SR was measured in the commercial membrane compared to the case of 4VBC*. A significantly lower degree of functionalisation has been measured in samples of 4VBC ($F_{4VBC}$= 21±3 mol.%) in comparison to samples of 4VBC-MA ($F_{4VBC-MA}$= 97±2 mol.%) [30,31]. The above observation therefore provides indirect evidence of the impact of the specific functionalisation route (i.e. single vs. sequential) on the crosslink density aiming to accomplish UV-cured AC prototypes with controlled swelling properties and enzymatic degradability. Accordingly, sequentially functionalised samples of 4VBC-MA* ($F_{4VBC-MA}$= 97±2 mol.%) displayed nearly twice the mass ($\mu_{rel}$ = 50±4 wt.%) of 4VBC* ($F_{4VBC}$= 21±3 mol.%; $\mu_{rel}$ = 24±4 wt.%), following a 14-day collagenase incubation of both samples *in vitro* (37 °C, 1 CDU·ml$^{-1}$) [30]. Unlike the case of UV-cured samples, however, insignificant variations in *SR* were recorded in the samples of Bio-Gide® during the 24-hour incubation in each aqueous medium (Figure 3C-D). Given that Bio-Gide® is a bilayer matrix of decellularised porcine collagen [47,48], the retention of native fibrillar organisation of collagen, as well as the presence of an additional dense layer in the membrane structure, explain the relative insensitivity of this product to the water-induced swelling in the above aqueous media.

**3.3 Impact of the UV-cured network architecture on compression properties**

The mechanical competence of GBR membranes *in vivo* is one of the primary factors to impact their clinical performance. Specifically, the compressibility of the GBR membrane is key to achieving a secure edge closure during clinical fixation, ensuring minimal risks of membrane instability and sinking into the bone defect during the healing process [49,50]. Compression tests were therefore carried out on the membrane prototypes directly after UV curing, as well as following the ethanol series dehydration and equilibration in PBS, aiming to investigate whether the ethanol series-induced microstructure densification observed by SEM (Figure 2) was reflected in any change in compression properties at the macroscale. PBS was selected as a near-physiologic aqueous environment for sample rehydration, and also in light of the intermediate swelling profiles obtained in this medium with respect to the case of distilled water or saline solution (Figure 3). On the other hand, no measurements could be carried out on Bio-Gide®, given its significantly smaller height ($h$= 0.44±0.11 mm) [51] compared to the samples of either F-4VBC* *($h$= 4.13±0.76 mm)* and F-4VBC-MA* ($h$= 6.19±1.35 mm) (Table 1) or R-4VBC* ($h$= 2.87±0.35 mm) and R-4VBC-MA* ($h$= 5.04±0.71 mm).

The resulting stress-compression curves (Figure 4) revealed a *J*-shaped mechanical response, reflecting the chemical composition of biological tissues [30,31,52,56,57]. A predominately elastic behaviour was observed up to the point of sample breakage in both hydrogels 4VBC* (Figure 4A-B) and 4VBC-MA* (Figure 4C-D), in both the freshly synthesised state or following ethanol series dehydration and rehydration in PBS. At the collapse plateau regime [40], i.e. at 25% of compression, lower values of stress were recorded with samples F-4VBC* *($\sigma_{25}$= 5±1 kPa, Figure 4A)*, compared to the variants R-4VBC* obtained via ethanol series dehydration and PBS equilibration ($\sigma_{25}$= 3±1 kPa, Figure 4B). On the other hand, comparable values were revealed by the sequentially functionalised networks of 4VBC-MA* in both freshly synthesised state and dehydrated and rehydrated state ($\sigma_{25}$= 2±6–6±6 kPa, Figure 4C-D). These observations suggest that the presence of pores and pore walls in the freshly synthesised samples F-4VBC* provides an interface through which mechanical damage and crack propagation can be delayed [53].

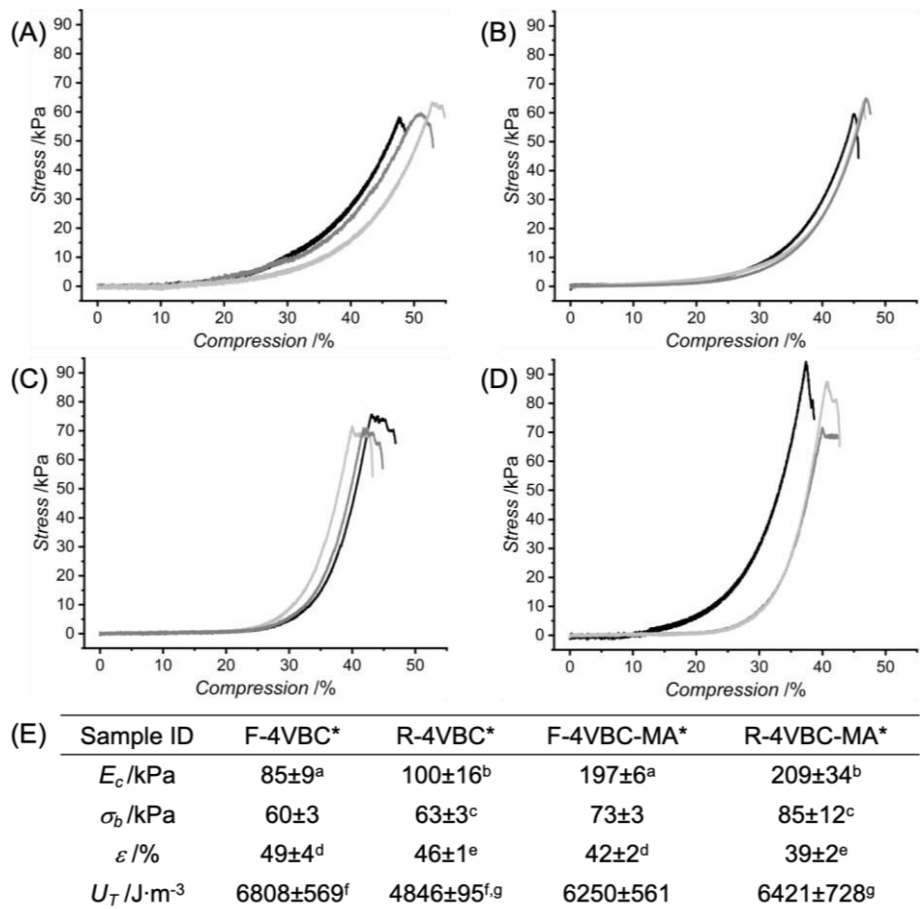

**Figure 4.** Stress-compression curves (n=3) of UV-cured collagen samples in the freshly synthesised state (A, C) and following ethanol series dehydration and equilibration in PBS (B, D). (A): F-4VBC*; (B): R-4VBC*; (C): F-4VBC-MA*; (D): R-4VBC-MA*. (E): Extracted data of compression modulus ($E_c$), stress at break ($\sigma_b$), compression at break ($\varepsilon$) and toughness ($U_T$). Data are presented as mean ± standard deviation. Same superscripts indicate statistically significant means ($p< 0.05$, Bonferroni test).

This hypothesis is supported by measurements of toughness ($U_T$), showing significantly higher values of $U_T$ in samples F-4VBC* compared to the corresponding rehydrated variants R-4VBC* (Figure 4E). The comparable values of stress measured with both F-4VBC-MA* and R-4VBC-MA* in the same collapse plateau regime (25% compression) also agree with the corresponding insignificantly different variation in toughness (Figure 4E). The above trends in $\sigma_{25}$ and $U_T$ likely reflect the significantly higher degree of functionalisation and crosslink density in samples of 4VBC-MA* compared to the 4VBC* variants [30,31,54], and an irregular distribution of micropores within the material.

Insignificantly higher compression modulus ($E_c$) and stress at break ($\sigma_b$), together with an insignificantly lower elongation at break ($\varepsilon$), were recorded with the dehydrated and rehydrated samples of either R-4VBC* or R-4VBC-MA* with respect to the corresponding freshly synthesised hydrogels. These observations support the microstructure densification induced by the ethanol series sample dehydration, as previously evidenced by the reduction in membrane height and the suppression of internal micropores (Figure 2). Therefore, the presented ethanol series dehydration process provides an additional dimension to adjust the compression strength and minimise risks of damage during fixation *in vivo*.

Other than the post-synthesis treatment, the UV-cured sequentially functionalised samples revealed ca. two-fold increase in $E_c$ and ca. 15% decrease of $\varepsilon$ with respect to the single functionalised variants, in both the fresh and rehydrated states (Figure 4E). Furthermore, a significantly higher $U_T$ was measured in samples R-4VBC-MA* compared to R-4VBC*. The decreased degree of AC functionalisation and crosslink density in the single-functionalised crosslinked network (4VBC*) compared to the sequentially functionalised variant (4VBC-MA*) agree with the above trends in mechanical properties, in line with previous reports [27,30,31,54]. Most importantly, the sequentially functionalised network architecture proved to equip the dehydrated and rehydrated membrane R-4VBC-MA* with significantly higher compression strength and toughness compared to membrane R-4VBC*, which is ideal to ensure mechanical integrity during clinical handling and surgery.

The mechanical response of crosslinked collagen is dominated by the crosslink type and density and includes a linear regime characteristic of molecules gliding and a second stiffer elastic regime related to the stretching or densification of the collagen molecules [40,55]. The aforementioned compression results therefore confirm the major role played by the molecular architecture of the UV-cured network of AC molecules, whereby either single- or sequential functionalisation of AC ensures a predominantly elastic behaviour prior to sample break. This compression behaviour is desirable in GBR therapy, as it aims to prevent mechanical failure *in vivo* and ensure the preserved volume of the CSBD following membrane implantation.

### 3.4 Local elasticity determined via AFM nanoindentation

AFM nanoindentation was employed to quantify the $E_{AFM}$ and the $F_A$ across the UV-cured samples R-4VBC* and R-4VBC-MA*, given the key role of these mechanical characteristics in ensuring successful GBR therapy. The $E_{AFM}$ was quantified by nanoindentation to determine its spatial variation at the nano-to-microscale and to compare these with values extracted from the elastic region of the macroscopic compression tests.

Five 10×10 µm regions were indented in the central region of each sample, to gain information on the distribution of mechanical properties across the sample surface. The relatively low spatial resolution employed (1 µm) was selected to facilitate higher throughput and reliable statistical comparison between the samples. Force-indentation curves (FC) from the 10×10 FV maps showed the typical response of an elastic half-space material (Figure 5). The $F_A$ measured between the carbon bead and both the AC membranes and the agarose controls was typically low compared to the maximum indentation force. Therefore, the loading force traces were fitted to the Hertz model for a spherical indenter on a plane to determine the $E_{AFM}$. The R-4VBC-MA* gave the highest values of $F_A$ of all samples (Figure 5B and Figure S4, Supp. Inf.), with the highest values representing ~20% of the maximum loading force.

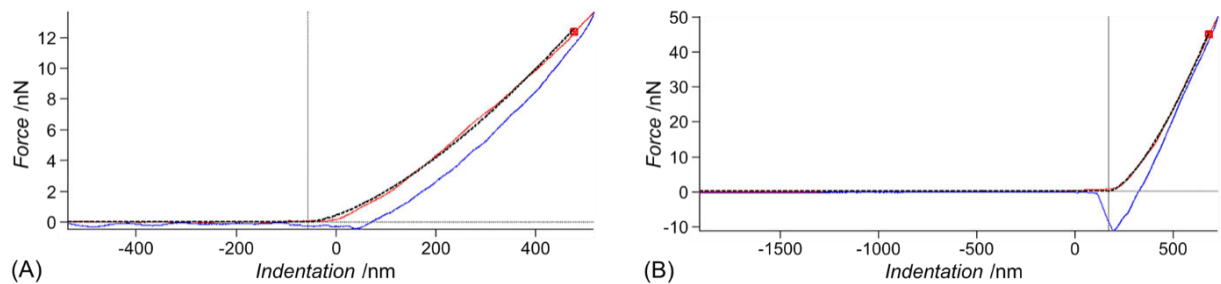

**Figure 5.** Exemplary AFM force-indentation curves recorded on PBS-rehydrated samples of R-4VBC* (A) and R-4VBC-MA* (B). The dashed black line depicts Hertz contact mechanics model fitting of the loading force (red).

For every sample, there were no statistically significant differences found between the five different areas tested, and hence all 500 FCs for each were collated, yielding non-normal distributions of $F_A$ and $E_{AFM}$ (Figure S5, Supp. Inf.). The range of concentrations of the hydrated agarose gel controls was selected to span the expected range of the AC membranes based on the measured bulk moduli. The median values of $E_{AFM}$ in the agarose control increased in a non-linear fashion (Figure S6, Supp. Inf.) from ~1 kPa (in the 3 wt.% agarose control) to >200 kPa (in the 12 wt.% agarose control), in line with the decreased porosity observed by SEM (Figure S3, Supp. Inf.). Consequently, these control experiments validated the AFM indentation approach, whereby the micron-sized spherical bead probe tip ($R$= 1500±10 nm) was selected to ensure a reproducible contact area during nanoindentation, minimising any surface irregularities [56,57]. This was key to offset any measurement variability due to the effects of internal micropore structure, and to mitigate potential nano-to-microscale heterogeneity in the gel samples. The contact area (A) at typical maximum loading force was ~4 µm$^2$, mapping the properties between the nano and micro length scales (i.e. 100 nm < $\sqrt{A}$ < 1 µm), such that the lateral length scale at maximum load was ~2 µm.

The $F_A$ at the sample surfaces was determined from the maximum negative value of force upon retraction/unloading of the bead after indentation (Figure 5). The sample of R-4VBC* exhibited negligible median values ($F_A$< 1 nN), as did the agarose control ($F_A$< 1 nN), whereas R-4VBC-MA* showed a statistically significant increase in the range of ~2 to ~49 nN (Figure 6), with a median value of 10 nN.

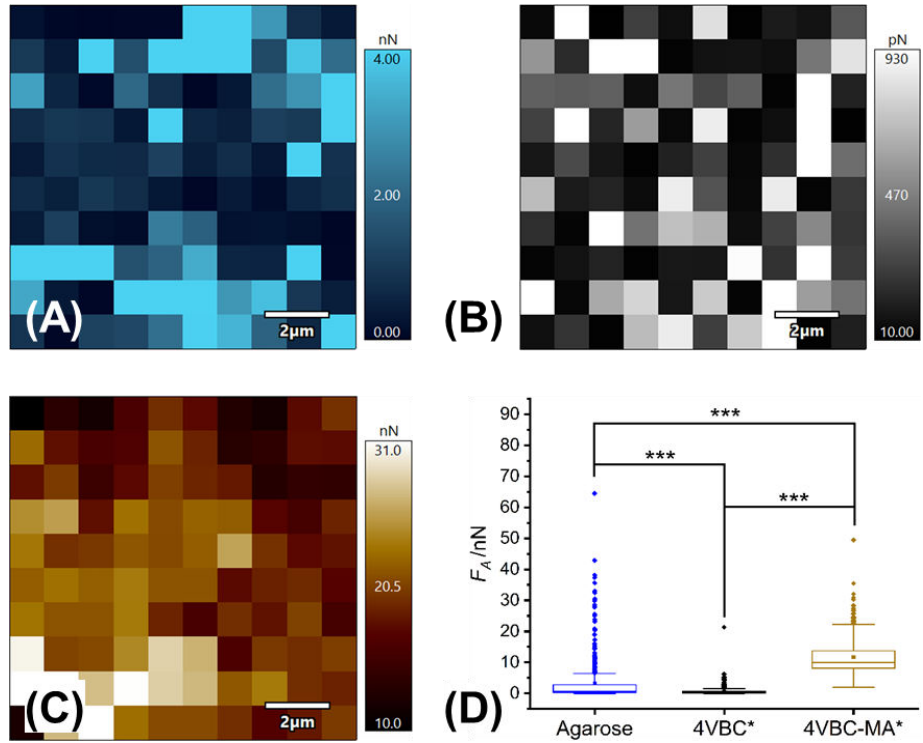

**Figure 6.** Determination of adhesion force ($F_A$) via AFM. (A-C): Representative $F_A$ maps measured in the 3 wt.% Agarose control (A) and PBS-rehydrated samples of R-4VBC* (B) and R-4VBC-MA* (C). (D): Box plots of $F_A$ data extracted from five adhesion maps of Agarose control (n= 441) and samples of R-4VBC* and R-4VBC-MA* (n= 500). (□): IQR; (■): mean; (—): median; (I): 1.5×IQR; (♦): outlier. *** $p$ <0.0001 (Kruskal–Wallis test).

More than one order of magnitude decrease in $F_A$ were therefore recorded in Agarose (Figure 6A) and 4VBC* (Figure 6B) compared to 4VBC-MA* (Figure 6C), whereby averaging all single map values generated median (IQR) adhesion forces of 0.7 (2.5) nN, 0.2 (0.6) nN and 10 (6) nN, respectively (Figure 6D). These results therefore support the direct relationship between the crosslink density of the molecular network and the macroscopic adhesivity [58, 59, 60]. Here, the development of π–π interactions between the 4VBC-crosslinking residues and the probe may provide an additional contribution to the overall adhesion force, as previously observed via chemical force microscopy between a benzylated AFM probe and a benzylated substrate [61].

Typical FV maps revealed spatial variations of $E_{AFM}$ in the agarose control (Figure 7A) and both the AC samples (Figure 7B-C), indicating microscale heterogeneities in both types of network [62].

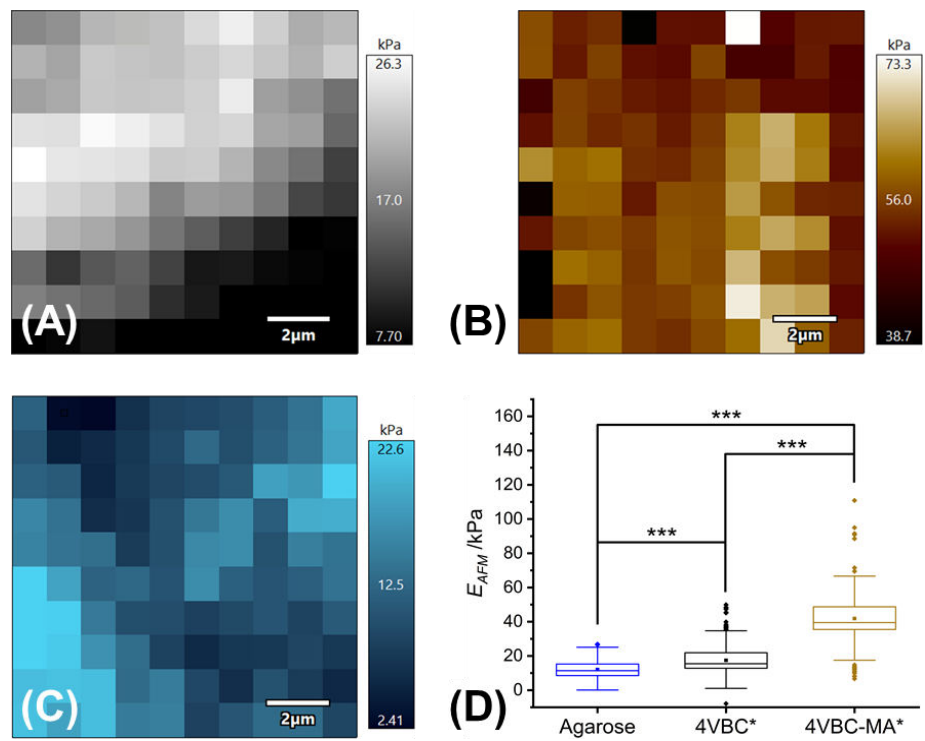

**Figure 7.** Determination of elastic modulus ($E_{AFM}$) via AFM. (A-C): Representative $E_{AFM}$ maps measured in the 3 wt.% agarose control (A) and PBS-rehydrated samples of R-4VBC* (B) and R-4VBC-MA* (C). (D): Box plots of $E_{AFM}$ data extracted from five modulus maps of Agarose control (n= 500) and samples of 4VBC* and 4VBC-MA* (n= 498). (□): IQR; (■): mean; (—): median; (I): 1.5×IQR; (♦): outlier. *** $p$ <0.0001 (Kruskal–Wallis test).

Averaging all values of $E_{AFM}$ determined in each indentation curve across all FVMs revealed significant differences between the two UV-cured membranes (Figure 7D, $p$ <0.0001). A nearly two-fold increase in $E_{AFM}$, i.e. median (IQR), was measured with R-4VBC-MA* ($E_{AFM}$= 40 (13) kPa) with respect to R-4VBC* ($E_{AFM}$= 15 (9) kPa, $p$ <0.0001), whereby the 3 wt.% agarose control ($E_{AFM}$= 11 (7) kPa, $p$ <0.0001) was selected for comparison from the full range of controls (Figure S6, Supp. Inf.).

Similar trends were observed for the AC membranes through the bulk compression tests in the 20-35% deformation range (Figure 4E), whereby R-4VBC-MA* revealed higher compression modulus ($E_c$= 209±34 kPa) compared to R-4VBC* ($E_c$= 100±16 kPa), in the dehydrated and rehydrated state that was tested during AFM.

The surface topography can be extracted from the FV maps at a given load to understand any correlation between structure and mechanics using force detection by AFM microbead nanoindentation (Figure S4, Supp. Inf.). To ensure that there was no influence of the local surface topography of the samples (both AC membranes and agarose gels) on either the adhesion or modulus maps, image cross-correlation of each map with its corresponding extracted height map (at maximum load) was carried out. The majority of cross-correlation values ($r_{xy}$ >95% for $F_A$; $r_{xy}$ ~70% for $E_{AFM}$) was between +/-0.5, indicating no significant correlation between mechanical properties and surface microstructure of the samples.

The $E_{AFM}$ values of the dehydrated and rehydrated AC membranes are approximately five-fold lower than the bulk modulus values, which we attribute to length scale dependency of probing the networks at the nano-to-microscale compared to the macroscale. The trend in $E_{AFM}$ between the single- and sequentially functionalised is the same as the one observed in $E_c$, however, with the former also showing a two-fold increase from R-4VBC* to R-4VBC-*MA**.

Overall, the AFM nanoindentation results therefore provide further evidence of the major role played by the molecular architecture, in terms of degree of AC functionalisation and consequent crosslink density, on the mechanical behaviour of the collagen networks across multiple length scales, in line with previous investigations [28]. The increased adhesivity exhibited by samples R-4VBC-MA* is relevant for GBR therapy, aiming to equip the dental membrane with space maintenance capability. This characteristic may be exploited to overcome the need for fixation *in situ*, which is typically achieved by suturing. Indentation measurements using a protein-coated bead could be useful to confirm the sample adhesivity with a simple tissue mimetic.

### 3.5 Suture-holding properties of UV-cured AC membranes

Suturing of the dental membrane to the defect site is typically required to ensure minimal movements of the membrane, preserve the defect volume, and prevent the growth of soft tissue during the bone regeneration process. While the use of titanium pins is an ideal fixation choice [63], due to their excellent biocompatibility, small size, and ease of manipulation, they are relatively expensive compared to traditional fixation methods and require surgical removal at later stages, raising unnecessary patient safety risks. On the other hand, sutures are widely used to close an injury or a cut within a soft tissue, or to firmly connect an implant to the surrounding tissues, following e.g. vascular [64], orthopaedic [65] or dental [66] trauma. Given its cost-effectiveness, suture-mediated membrane fixation was adopted in this study to evaluate the suture-holding properties of the two UV-cured AC prototypes, alongside Bio-guide®. Tensile measurements were conducted on sutured membranes (Figure 8) in the dry state (i.e. with 4VBC* and 4VBC-MA*) and following 1-hour and 24-hour sample incubation in PBS, to quantify the mechanical performance in near-physiological conditions.

Exemplary stress-strain curves presented varied mechanical responses depending on both the sample group and the state of hydration (Figure 9A-C). The UV-cured samples exhibited an initial break in the dry state (*displacement*: 5-7 mm) and after 1 hour in PBS (*displacement*: 5-8 mm), prior to complete pull-out, which was recorded at a displacement of ca. 9 mm (Figure 9A-B). At the initial break, significantly higher values of *BSS* were observed with 4VBC* (*BSS*: 19±1–20±5 N·mm$^{-1}$) with respect to 4VBC-MA* (*BSS*: 7±3–13±2 N·mm$^{-1}$), while the latter samples revealed a significant reduction in *BSS* even after 1 hour in PBS, compared to the dry state.

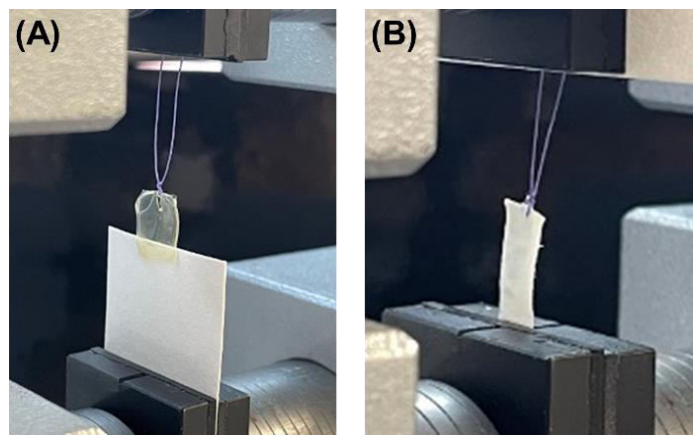

**Figure 8.** Photographs of exemplary sutured samples of UV-cured collagen-based dental membrane (A) and Bio-Gide® (B) following 24-hour incubation in PBS (10 mM, pH 7.4) and clamping onto the tensile testing apparatus. The pull-out test was conducted with a tensile rate of 0.02 mm·s$^{-1}$, proceeding until the suture was fully pulled out.

While comparable pull-out responses were measured between the dry and 1-hour incubated samples of 4VBC*, lower forces were obtained when dry samples of 4VBC-MA* were incubated in PBS for 1 hour prior to testing (Figure 9A-B and Figure S7, Supp. Inf.). This observation indicates a slower water-induced plasticisation in the former compared to the latter sample and reflects the major role played by the molecular architecture of the UV-cured AC network on the wet state mechanical properties of resulting membrane. In 4VBC*, $\pi$-$\pi$ aromatic interactions are expected between the 4VBC-crosslinked molecules of AC, due to the functionalisation of AC with the benzyl residue of 4VBC. Such aromatic interactions are less likely to take place in 4VBC-MA* due to the higher functionalisation of AC with MA, rather than 4VBC, residues ($F_{4VBC-MA}$= 97±2 mol.%; $F_{4VBC}$= 21±3 mol.%) [30], and the UV-induced synthesis of methacrylated, in addition to aromatic, crosslinking segments. Consequently, while 1-hour incubation in PBS led to a detectable water-induced plasticisation in 4VBC-MA*, only, stark changes in pull-out response were measured following longer incubation time, i.e. 24 hours (Figure 9C). A dominant elastic response was observed in all samples, as evidenced by the suppression of the initial break. While comparable values of pull-out displacement (~8.5 mm) were measured, a significantly lower *SRS* was recorded in both R-4VBC* and R-4VBC-MA* with respect to the corresponding dry and 1-hour incubated samples, in line with the water-induced plasticisation of the crosslinked AC molecules after 24 hours in PBS [66,67].

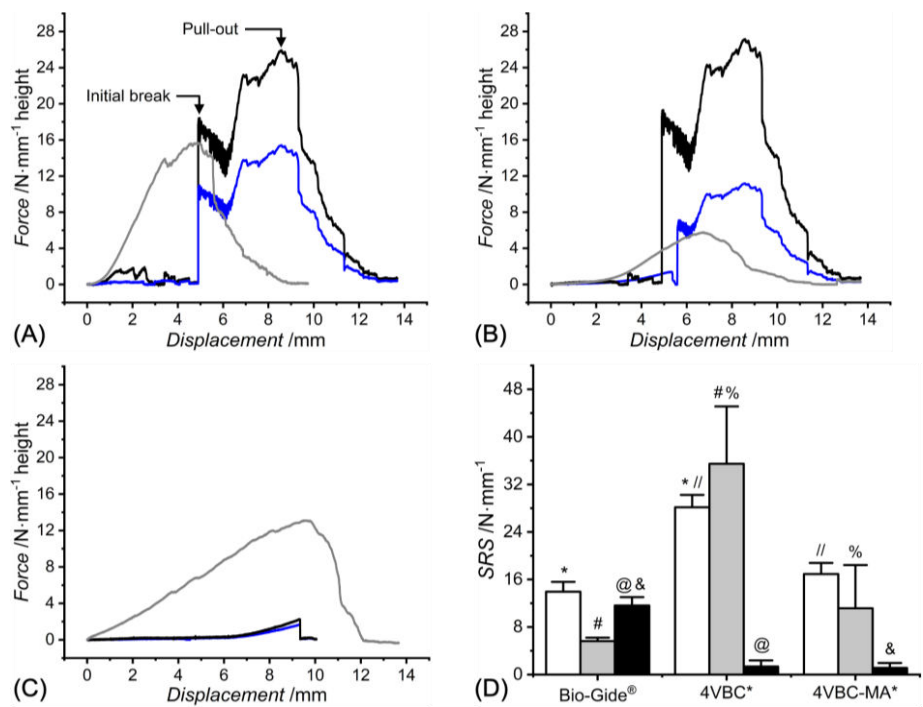

**Figure 9.** Suture pull-out measurements (n=3) on Bio-Gide® (grey), 4VBC* (black), and 4VBC-MA* (blue). (A-C): representative force-displacement curves recorded in the dry state (A) and following sample incubation in PBS for 1 hour (B) and 24 hours (C). Values of force are normalised by the sample height. The force measured at either the pull-out or the initial break corresponds to the suture retention strength (SRS) and break starting strength (BSS), respectively. (D): Values of SRS recorded on dry samples (white column) and following incubation in PBS for 1 hour (grey column) and 24 hours (black column). *, #, @, &, %: $p < 0.05$ (Bonferroni test).

Analysis of the force-displacement curves obtained with sutured samples of Bio-Gide® indicated a different tensile response with respect to the dry and 1-hour-incubated UV-cured prototypes, whereby no initial break and insignificantly different pull-out displacements (5–26 mm) were recorded (Figure 9C). Similar profiles were observed following 24-hour incubation of Bio-Gide® in PBS, whereby a somewhat lower SRS was measured consequent to the uptake of water in the membrane (Figure 3C-D). The more uniform mechanical response revealed by Bio-Gide® with respect to the UV-cured samples, even in the dry state, is attributed to the fibrillar organisation of collagen molecules, the lack of covalent crosslinks, and the presence of the bilayer microstructure.

The water-induced plasticisation observed in 4VBC* appeared to be slower than Bio-Gide® and detectable only after 24 hours in PBS, at which point comparable tensile profiles were recorded, though with significantly higher values of SRS revealed by Bio-Gide® with respect to the two AC membranes. This latter observation is attributed to the significantly lower values of SR exhibited by the Bio-Gide® (Figure 3), and the difference in microstructure. In contrast to the UV-cured samples, Bio-Gide® is a non-crosslinked, decellularised membrane, whose mechanical response is primarily ruled by the presence of the bilayer microstructure, the material density in each layer and the structural (i.e. fibrillated) organisation of collagen molecules [33].

A comparison of the SRS values measured in all samples confirmed the significant effect of the hydration state on membrane suturing ability (Figure 9D). Samples of 4VBC* displayed the highest values in both the dry state (SRS= 28±2 N·mm$^{-1}$) and following 1 hour in PBS (SRS= 35±10 N·mm$^{-1}$), followed by 4VBC-MA* (SRS: 11±7–17±2 N·mm$^{-1}$) and Bio-Gide® (SRS: 6±1–14±2 N·mm$^{1}$). This trend was reversed after 24 hours in PBS, whereby the commercial membrane revealed significantly higher suture-holding properties (SRS= 12±1 N·mm$^{-1}$), whilst comparable, but significantly lower values were recorded in both UV-cured AC prototypes (SRS= 1±1 N·mm$^{-1}$), in line with previous differences in swelling profiles (Figure 3).

On the one hand, the results obtained in the dry state and following 1 hour in PBS confirm that the dehydration-induced microstructure densification in the UV-cured samples (Figure 2) represents a suitable route to accomplish competitive clinical handling capability. On the other hand, these results show that the crosslinked architecture of the UV-cured network of 4VBC* enables significantly increased *SRS* with respect to Bio-Gide®, despite the more than 5-fold increase in water uptake measured in the former samples after 1 hour in PBS (Figure 3). Counterintuitively, the value of *SRS* measured in 4VBC* proved to be higher than the one of 4VBC-MA*, despite the significantly lower degree of functionalisation (and consequent crosslink density) [30,31], and the significantly decreased values of $E_c$ and $U_T$, measured in the former compared to the latter samples. Considering the different mechanical responses of the samples in tensile and compressive deformation, a possible explanation for this observation is that the ethanol-induced dehydration of 4VBC* leads to aromatic bulky clusters and an increase in free volume, consequent to the development of $\pi-\pi$ interactions between crosslinked AC molecules. These additional physical crosslinks are unlikely to take place in the samples of 4VBC-MA*, due to the significantly higher molar content of methacrylate crosslinking segments [30]. The corresponding increase in free volume is expected to facilitate the tearing of the sample due to the force from the suture [64,68].

**Conclusions**

Two UV-cured prototypes were made with either single-functionalised (4VBC*) or sequentially functionalised AC (4VBC-MA*). Both porous and pore-free microstructures were obtained independently of the crosslinked architecture introduced at the molecular scale so that enhanced barrier functionality was evidenced following ethanol series dehydration and consequent merging of internal pores. Samples of 4VBC* highlighted the highest *SR* in either distilled water, PBS or saline solution, while comparable variations between aqueous environments were observed in both crosslinked architectures. Both the sequential functionalisation of AC in the UV-cured network and the ethanol series-induced material densification resulted in increased compression modulus, compression strength and toughness. AFM nanoindentation yielded the same trend in elastic modulus for the two architectures at the micro-/nanoscale compared to the compression tests at the macroscale, and importantly revealed that 4VBC-MA* has enhanced surface adhesivity. Suture pull-out tests revealed an increased suture retention strength with both dry and 1-hour-incubated samples of 4VBC* compared to the case of 4VBC-MA* and Bio-Gide®, indicating a slower water-induced plasticisation of the former compared to the latter sample. Longer incubation time induced a significant decrease in suture retention strength in both UV-cured samples compared to the Bio-Gide®, reflecting the relatively low swelling ratio and the presence of a bilayer microstructure in the latter samples. This study indicates the major role played by the crosslinked architecture as well as structural and molecular organisation on the compression properties, adhesivity, and suturing ability of the UV-cured membrane. These characteristics could be further exploited to accomplish independent adjustment of material properties for competitive GBR functionality. Further investigations will be key to establish correlations between these characteristics and controlled membrane biodegradability for predictable GBR outcomes.

**Data access statement**

All data that support the findings of this study are included within the article (and any supplementary files).

**Acknowledgements**

This study was partially supported by the Royal Society International Exchanges Cost Share Award, UK (IEC\NSFC\223289). The authors acknowledge technical assistance from Lekshmi Kailas supported by EPSRC Strategic Equipment Scheme grant (EP/R043337/1) within the University of Leeds AFM facility. Michael Brookes is gratefully acknowledged for his technical assistance with scanning electron microscopy.

**Conflict of interest statement**

D.W. and G.T. are named inventors on a patent related to the fabrication of collagen-based materials. They have equity in and serve on the board of directors of HYFACOL Limited.

**Supporting Information**

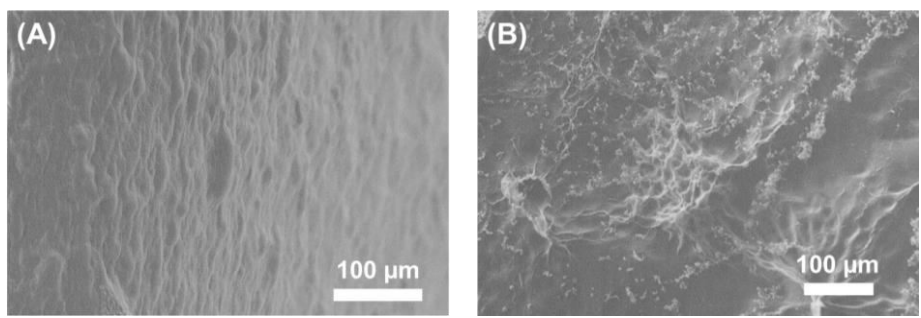

**Figure S1.** Higher magnification SEM images of freshly synthesised samples of F-4VBC* (A) and F-4VBC-MA* (B).

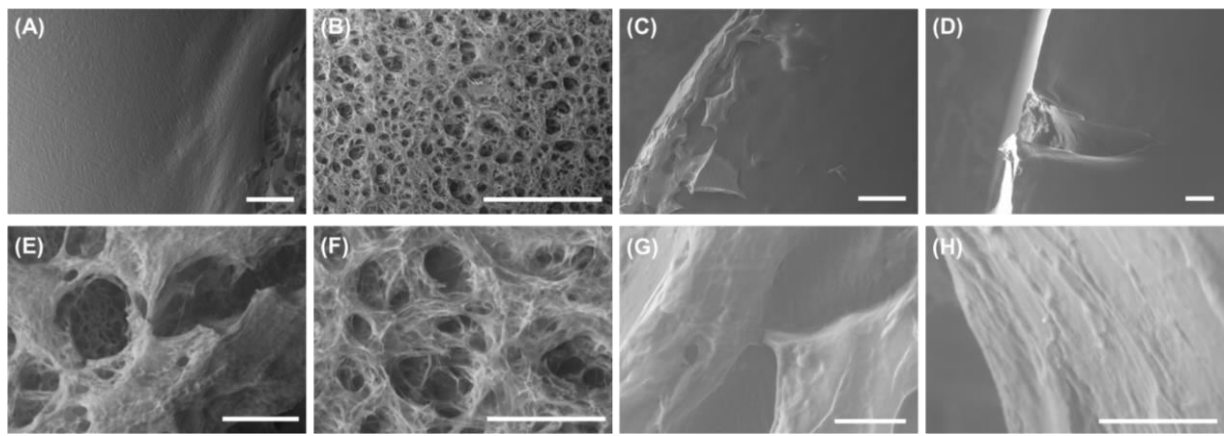

**Figure S2.** Cool-stage SEM images captured at low (A-D, scale bar ≈200 μm) and high (E-H, scale bar ≈50 μm) magnification of fresh agarose controls prepared with varied agarose concentrations. (A, E): 1.5 wt.% agarose; (B, F): 3 wt.% agarose; (C, G): 6 wt.% agarose; (D, H): 12 wt.% agarose.

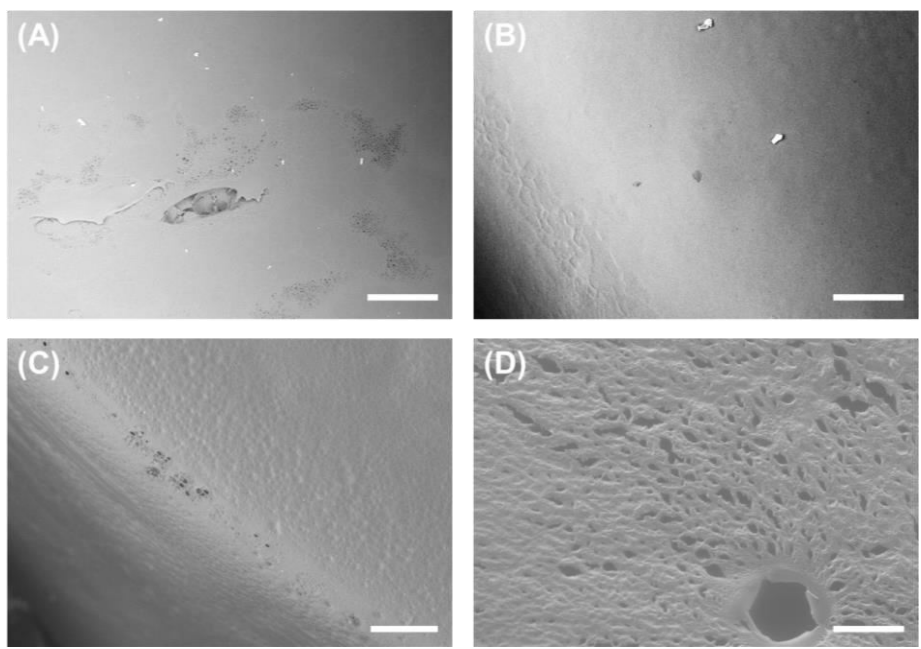

**Figure S3.** Cool-stage SEM images of dehydrated agarose controls prepared with varied agarose concentrations. (A): 1.5 wt.% agarose; (B): 3 wt.% agarose; (C): 6 wt.% agarose; (D): 12 wt.% agarose. Scale bar ≈ 200 µm.

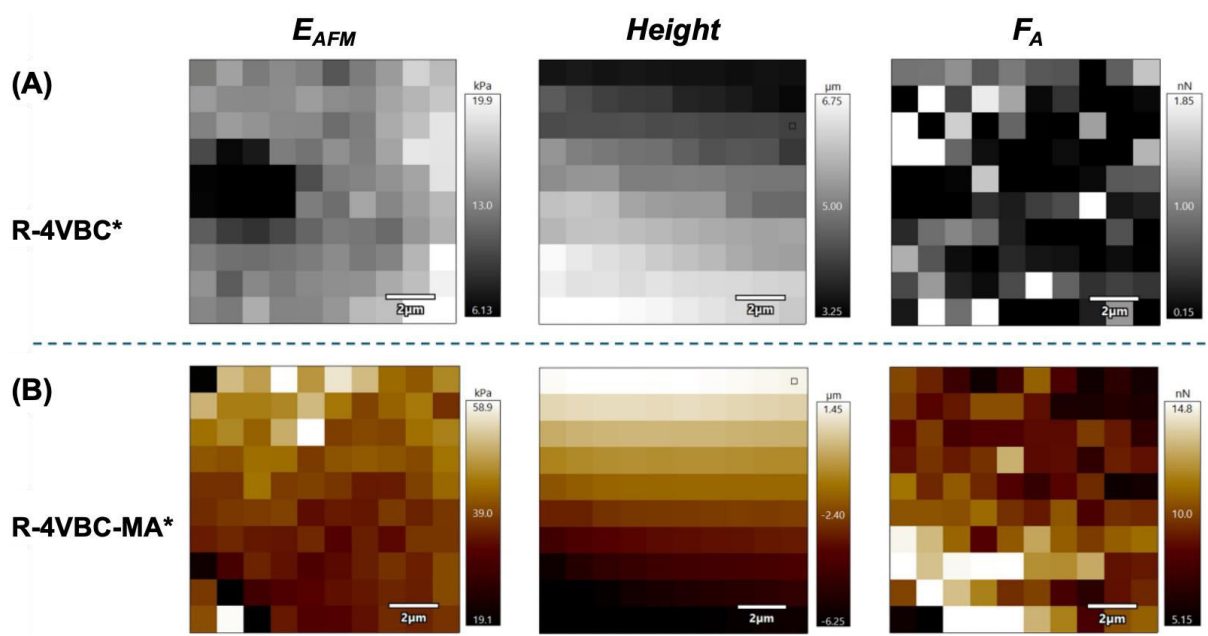

**Figure S4.** Example Force-Volume (FV) Maps from one of the five 10x10 µm regions of the PBS-rehydrated membranes of R-4VBC* (A) and R-4VBC-MA* (B) prepared with varied crosslinked architecture. The *Height* map represents the surface topography of the sample at maximum load, where a slope may be detected, but this can also be affected by instrumental drift during the FV measurements. Visually one observes that neither the elastic modulus ($E_{AFM}$) nor the adhesion force ($F_A$) is correlated to the *Height*, or to each other, at the microscale. This was confirmed by the image cross-correlation coefficients for all five regions tested on each sample, as determined in ImageJ (v.1.54).

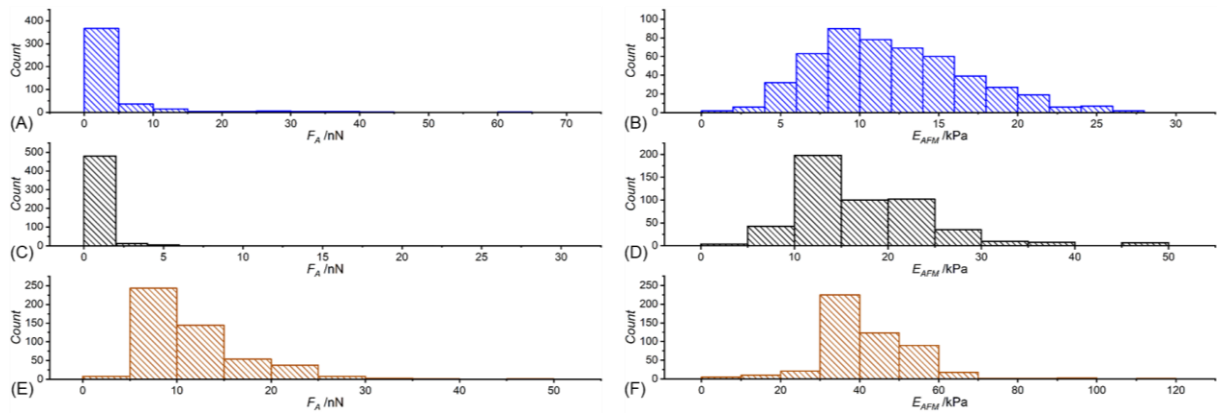

**Figure S5.** AFM distribution plots of adhesion force ($F_A$) and elastic modulus ($E_{AFM}$) measured on the agarose control (A, n= 441; B, n= 500) and samples of R-4VBC* (C, n =500; D, n= 498) and R-4VBC-MA* (E, n= 500; F, n= 498). N< 500 indicates that some force-indentation curves were discarded due to limited quality.

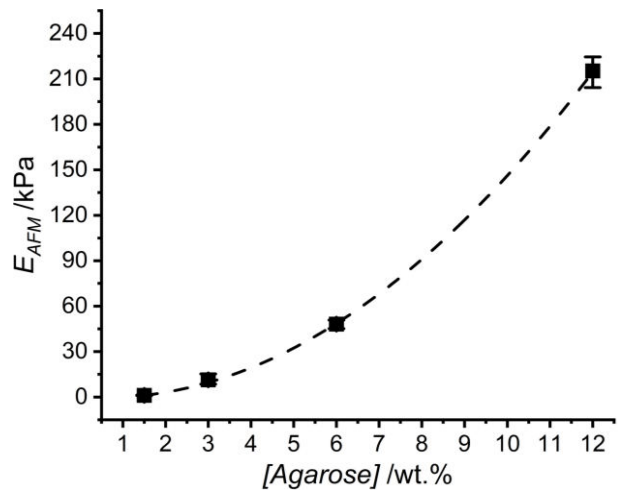

**Figure S6.** Variation of elastic modulus ($E_{AFM}$) determined by microbead AFM in agarose controls prepared from solutions with increasing agarose concentration. The dashed line is the polynomial fit of the median data points (y= $1.6309x^2$-1.6899x-0.0023; $R^2$= 0.99971). Data (n= 500) are presented as median (IQR).

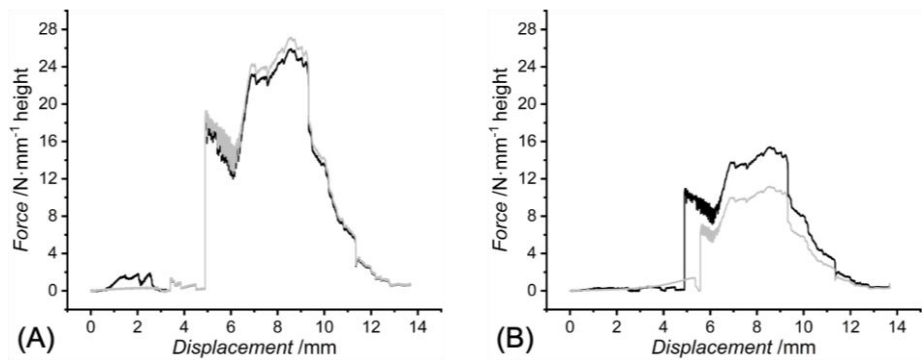

**Figure S7.** Force-displacement curves revealed by sample 4VBC* (A) and 4VBC-MA* (B) during suture pull-out measurements in the dry state (black) and following 1-hour incubation in PBS (light grey).